\documentclass[preprint,showpacs,preprintnumbers,endfloats*,amsmath,amssymb,linenumbers,prb]{revtex4-1}
\usepackage{graphicx}
\usepackage{dcolumn}
\usepackage{bm}
\newcommand{\vecvar}[1]{\mbox{\boldmath$#1$}}

\begin{document}

\preprint{PRESAT-8501}

\title{First-principles study on atomic configuration of electron-beam irradiated C$_{60}$ clusters}

\author{Tomoya Ono}
\affiliation{Graduate School of Engineering, Osaka University, Suita, Osaka 565-0871, Japan}
\author{Shigeru Tsukamoto}
\affiliation{Peter Gr\"{u}nberg Institut \& Institute for Advanced Simulation, Forschungszentrum J\"{u}lich and JARA, D-52425 J\"{u}lich, Germany}

\date{\today}

\begin{abstract}
A theoretical study proposes the atomic configuration of electron-beam irradiated C$_{60}$ thin films. We examined the electronic structure and electron-transport properties of the C$_{60}$ clusters using density-functional calculations and found that a rhombohedral C$_{60}$ polymer with $sp^3$-bonded dumbbell-shaped connections at the molecule junction is a semiconductor with a narrow band gap while the polymer changes to exhibit metallic behavior by forming $sp^2$-bonded peanut-shaped connections. Conductance below the Fermi level increases and the peak of the conductance spectrum arising from the $t_{u1}$ states of a C$_{60}$ molecule becomes obscure after the connections are rearranged. The present rohmbohedral film, including the [2+2] four-membered rings and peanut-shaped connections, is a candidate to represent the structure of the metallic C$_{60}$ polymer at the initial stage of electron-beam irradiation.
\end{abstract}

\maketitle
\section{Introduction}
\label{sec:intro}
Intermolecular electron transport has attracted a great deal of attention in numerous fields of research accompanied by the progress in nanostructure-fabrication technology. Fullerene-based molecular crystals and films are of particular interest because of their very rich physical properties such as superconductivity and magnetism. The polymerization of C$_{60}$ molecules in the solid phase using various techniques, e.g., photoirradiation or electron-beam (EB) irradiation, have resulted in new forms of carbon materials.\cite{eklund} The first-principles study by Okada and Oshiyama reported that the band gap of the rhombhedral phase of the C$_{60}$ polymer linked by [2+2] four-membered rings, where two C atoms shared by the two adjacent hexagons in a C$_{60}$ molecule are covalently bonded to the C atoms shared by the two hexagons in the neighboring C$_{60}$ molecule (e.g., see Fig.~\ref{fig:1}), is significantly smaller than that of C$_{60}$ bulk.\cite{PhysRevB.68.235402} Onoe {\it et al.} found that C$_{60}$ films change to polymers accompanying the insulator-metal transition due to EB irradiation and that a one-dimensional (1D) peanut-shaped connection is formed, where C$_{60}$ molecules are linked by $sp^2$-like connections instead of $sp^3$-like connections,\cite{onoe} and intensive studies on electron transport in the C$_{60}$ polymer have been carried out up to now.\cite{onoeapl} First-principles calculations on the electronic structure of three-dimensional (3D) C$_{60}$ polymers have revealed that 1D peanut-shaped C$_{60}$ polymers are insulators and electrons are conducted across the 1D polymers via [2+2] four-membered rings between the two peanut-shaped C$_{60}$ polymers in the 3D hexagonal structure,\cite{ueda,PhysRevB.75.233410} which is a completely different configuration from the face-centered cubic structure of C$_{60}$ bulk.

Nakaya {\it et al.} recently revealed that the chemical linkage between C$_{60}$ molecules created by EB irradiation are thermally more stable than the [2+2] four-membered ring because the polymers are stable under thermal annealing at 220 $^\circ$C, which is higher than the decomposition temperature of the [2+2] four-membered rings, and they claimed that a peanut-shaped fullerene contributes to the metallic characteristics of EB irradiated film.\cite{nakaya} They also demonstrated by scanning-tunneling spectroscopy (STS) that the characteristic peak of spectra of the C$_{60}$ molecule disappears at its polymerized center after EB irradiation. Moreover, their scanning tunneling microscopy (STM) images demonstrated that the C$_{60}$ polymers retain a {\it rhombohedral} structure at the initial stage of EB irradiation; however, this result does not agree with the {\it hexagonal} structure derived by Onoe {\it et al.} using first-principles calculations.\cite{ueda,PhysRevB.75.233410} Thus, the physics underlying the generation of conductivity in EB irradiated C$_{60}$ films remains unclear.

In this paper, the atomic configuration of the EB irradiated C$_{60}$ film is proposed by using first-principles calculations. We examine what effect the bond configuration at the molecule junctions has on the electronic structure and electron-transport properties to interpret the generation of conductivity in EB irradiated C$_{60}$ films. Our findings are that the rhombohedral C$_{60}$ polymer consisting of the [2+2] four-membered rings and the $sp^3$-like interlayer connections create a semiconductor with a narrow band gap and the band gap vanishes when the $sp^2$-like interlayer connection is formed. We found significant differences between the conduction spectra of the dimers bonded by the $sp^3$-like and $sp^2$-like connections; the conductance of the $sp^2$-like bonded dimer was higher than that of the $sp^3$-like bonded dimer below the Fermi level and the peak of the conductance spectrum caused by the $t_{u1}$ orbitals of the $sp^3$-like bonded dimer was clear while that of the $sp^2$-like bonded dimer was obscure.

All calculations are performed within the framework of density functional theory\cite{dft} using the real-space finite-difference approach, which enables us to determine the self-consistent electronic ground state with a high degree of accuracy by using a timesaving double-grid technique.\cite{book,tsdg} The norm-conserving pseudopotentials\cite{norm} of Troullier and Martins\cite{tm} are used to describe electron-ion interaction, and exchange correlation effects are treated with local density approximation.\cite{lda}

\section{Electronic structure of C$_{60}$ polymers}
\label{sec:elect}
C$_{60}$ polymers form a triangular lattice with [2+2] four-membered rings in each polymerized layer and the layers are stacked along a direction perpendicular to the layers in a rhombohedral symmetry.\cite{PhysRevB.68.235402} When the {\it ABC}-stacking structure\cite{chen} is formed, the layers are bonded by $sp^3$-like connections, in which the hexagons on neighboring C$_{60}$ molecules are brought face to face. When the C-C bonds sheared by a hexagon and pentagon on the facing hexagons are distorted, the layers are linked by an $sp^2$-like connection consisting of hexagons and heptagons. After this, we will refer to $sp^3$-like interlayer connections as dumbbell-shaped and $sp^2$-like interlayer connection as peanut-shaped. The peanut-shaped connection is much stronger than the [2+2] four-membered ring because the energy to form it is smaller than that for the [2+2] four-membered ring by 0.02 eV/atom.\cite{tsukamoto} Therefore, the polymer can be reinforced by the peanut-shaped connection so that it can survive under thermal annealing. We first calculate the electronic structure of the C$_{60}$ polymers.

\begin{figure}
\includegraphics{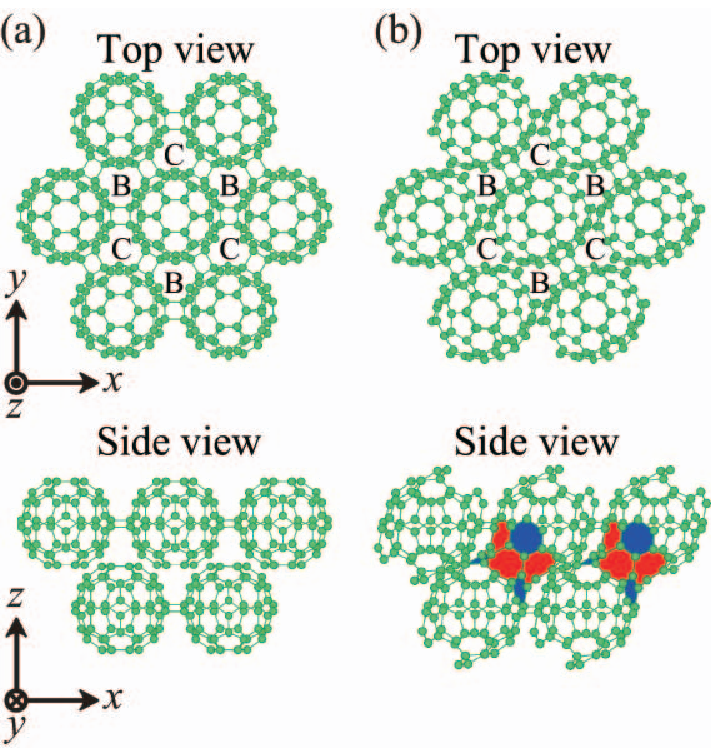}
\caption{(color online) Geometric structure of 3D C$_{60}$ polymer phases. (a) Polymer bonded by only [2+2] four-membered rings and dumbbell-shaped connections and (b) polymer in which one of three dumbbell-shaped connections deforms to the peanut-shaped connection. The shaded segments in (b) represent the hexagons (blue) and heptagons (red) of the peanut-shaped connection. B and C in the top views correspond to the positions of C$_{60}$ molecules in the upper and lower layers.}
\label{fig:1}
\end{figure}

\begin{figure}
\includegraphics{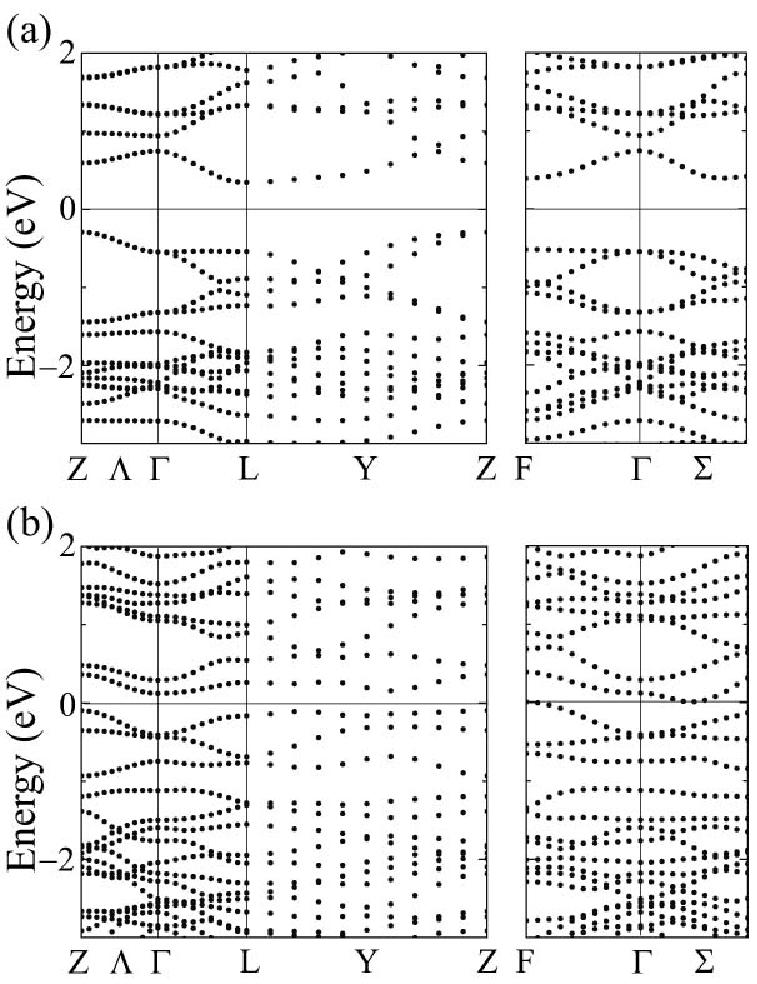}
\caption{Band structures of 3D C$_{60}$ polymers of models (a) and (b) in Fig.~\ref{fig:1}. Zero energy was chosen as the Fermi level.}
\label{fig:2}
\end{figure}

Figure~\ref{fig:1} shows two geometries of the models studied here. The C$_{60}$ polymers in model (a) are bonded by the [2+2] four-membered rings in the layer and linked by the dumbbell-shaped connections between the layers. In model (b), one of the dumbbell-shaped connections between the layers is deformed into the peanut-shaped connection. Since the metallic phase of the C$_{60}$ polymer is surrounded by the other phase of the C$_{60}$s at the initial stage of the EB irradiation according to the STM image,\cite{nakaya} its lattice parameters are not expected to be fully relaxed. We assume lattice parameters $a$=17.37 bohr, $c$=46.30 bohr and $\alpha=\pi/3$, which correspond to those for the dumbbell-shaped interlayer connection obtained by the x-ray diffraction pattern analysis,\cite{PhysRevLett.74.278} and employ these lattice parameters for both model (a) and model (b) to compare the contribution of the dumbbell-shaped and peanut-shaped connections to the electronic structures. Note that the stacking structure of the polymer presented in this study is not deformed from the rhombohedral bulk phase and corresponds to the experimentally observed STM image whereas the hexagonal polymers in Refs.~\onlinecite{ueda} and \onlinecite{PhysRevB.75.233410} require drastic structural deformation. Integration in the Brillouin zone is carried out using 24-point sampling and structural optimization for the atomic geometris is implemented for both models until the remaining forces for each atom are less than 36 meV/bohr. The calculated electronic band structures are depicted in Fig.~\ref{fig:2}. The energy band gap is reduced by forming peanut-shaped connections; the peanut-shaped polymer exhibits metallic behavior while model (a) is a semiconductor with a fundamental band gap of $\sim$ 0.6 eV. The formation energy of model (b) is larger than that of model (a) by 0.25 eV/atom although the formation energy of an isolated peanut-shaped dimer is smaller than that of a dumbbell-shaped dimer by 0.04 eV/atom. Thus, the peanut-shaped connections are not widely generated but C$_{60}$ molecules form small clusters with peanut-shaped connections in the film after EB irradiation.

\section{Transport properties of C$_{60}$ dimers}
\label{sec:trans}
The STS spectra, which have been employed to interpret the transport properties of molecules, show peaks which are distinctly associated with the electronic structure of molecules.\cite{th} However, to replicate the STS spectrum of the C$_{60}$ cluster in the C$_{60}$ film, the system including the large number of atoms are required because a C$_{60}$ molecule contains 60 atoms, which is not easy task by the present computational resources. Moreover, the number of peaks measured for the considered bias-voltage range have been smaller than the number of electronic states in the molecules within the corresponding energy window in some cases.\cite{antith} This fact implies that some states of molecules do not contribute to electron transport and that a simple interpretation of transport properties only in terms of DOS is insufficient. In addition, it is well known that the junction between C$_{60}$ molecules are a bottleneck for electron transport\cite{PhysRevLett.98.026804,PhysRevB.69.121408,PhysRevLett.103.206803} and the energy gap between the highest-occupied molecular orbital (HOMO) and the lowest-unoccupied molecular orbital (LUMO) decreases after polymerization.\cite{PhysRevB.75.233410,ueda,tsukamoto,nakayama} One of the present authors (S.T.) examined variations in the local density of states (LDOS) along the 1D C$_{60}$ polymer axis and demonstrated that the energy gap of LDOS at the molecule junction is larger than that in the C$_{60}$ molecules.\cite{tsukamoto} The $sp^2$-like connections of the peanut-shaped polymer might enhance electron transport because $\pi$ electrons are major carriers in metallic carbon nanotubes. Thus, it is of considerable interest to examine how the difference in the bond network between fullerenes contributes to electron transport through fullerenes.

\begin{figure}
\includegraphics{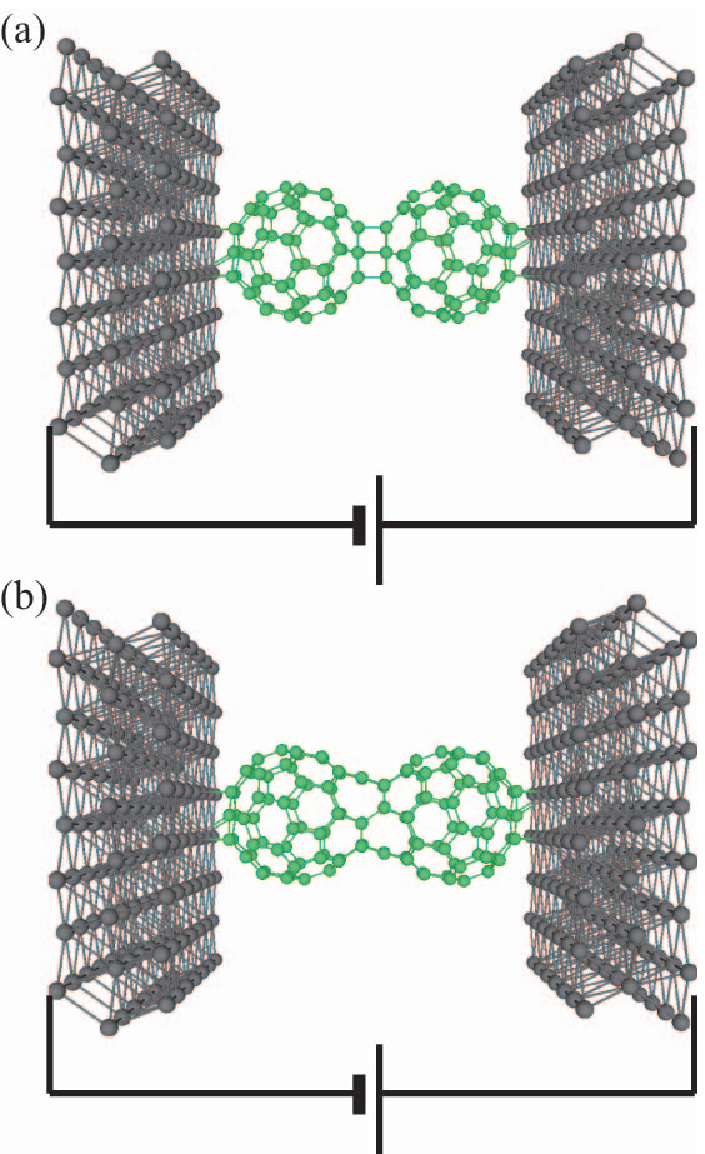}
\caption{(color online) Computational model where C$_{60}$ dimer is suspended between Al(111) electrodes. (a) Ddumbbell-shaped dimer and (b) peanut-shaped dimer.}
\label{fig:3}
\end{figure}

We explore the contribution of the peanut-shaped connection to electron-transport properties. Figure~\ref{fig:3} shows the computational model, where a C$_{60}$ dimer is sandwiched between electrodes. Since the lattice-constant mismatch between the C$_{60}$ polymers and the electrodes gives rise to further complex discussions on the transport properties, we employ the simplified models to focus on the difference in the contributions to the transport properties between the dumbbell-shaped and peanut-shaped connections. To determine the optimized atomic coordinates and Kohn-Sham effective potential, we use a conventional supercell under a periodic boundary condition in all directions with a real-space grid spacing of $\sim$ 0.32 bohr; the dimensions of the supercell are $L_x=37.87$ bohr, $L_y=37.49$ bohr, and $L_z=71.15$ bohr, where $L_x$ and $L_y$ are the lateral lengths of the supercell in the $x$- and $y$-directions parallel to the electrode surfaces, respectively, and $L_z$ is the length in the $z$ direction. Structural optimizations are implemented in advance for the isolated peanut-shaped dimer and it is then placed between the electrodes, where the three topmost surface atomic layers are atomistic Al(111) and the rest are aluminum jellium. The dimer is aligned at the hexagonal-close-packed-hollow site on the (111) surface facing a hexagon, which is the most stable configuration for a C$_{60}$ molecule on a face centered cubic (111) surface according to first-principles calculations.\cite{c60surface} The distance between the surface atomic layers on the left and right electrodes is set at 35.81 bohr so that the distance between the edge atoms of the dimer and the surface atomic layer of the electrodes corresponds to that reported by first-principles calculations. The dimer is relaxed between the electrodes. The atomic geometries of 36 carbon atoms for the dumbbell-shaped dimer at the molecule junction are modified from the above mentioned structure and structural optimization is also implemented.

We take a grid spacing of $\sim$ 0.47 a.u for the electron-transport calculations. We ensured that the decreased grid spacing and the enlarged supercell would not significantly affect our results. The scattering wave functions of the electrons propagating from the electrodes are determined by using the method of overbridging boundary matching.\cite{obm,kong,iobm} The retarded self-energy matrices for aluminum jellium are employed to include the rest of the semi-infinite electrodes. Since the DOS of aluminum is similar to that of free electrons, unfavorable effects from the DOS of the electrodes on the conductance spectra can be eliminated. We first calculate the Kohn-Sham effective potential using the supercell employed in structural optimization and then compute the scattering wave functions obtained non-self-consistently. It has been reported that this procedure is just as accurate in the linear response regime but significantly more efficient than performing computations self-consistently on a scattering-wave basis.\cite{kong2} The conductance of the dimers is described by the Landauer-B\"uttiker formula, $G=\text{Tr}(\textbf{T}^{\dag}\textbf{T})G_0$,\cite{buttiker} where $\textbf{T}$ is a transmission-coefficient matrix.

\begin{figure}
\includegraphics{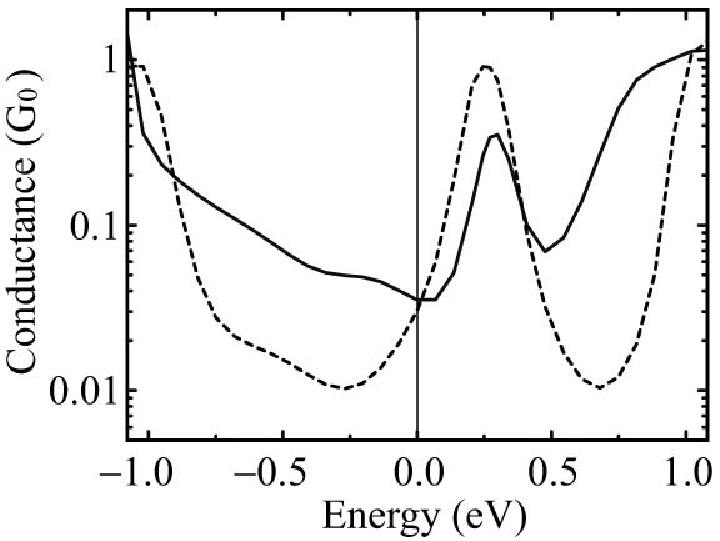}
\caption{Conductance spectra as function of energy of incident electrons. The dashed curve represents the conductance of the dumbbell-shaped dimer and the solid curve that of the peanut-shaped dimer. Zero energy was chosen as the Fermi level.}
\label{fig:4}
\end{figure}

Figure~\ref{fig:4} plots the conductance spectra of the dumbbell-shaped and peanut-shaped dimers as a function of the energy of incident electrons. The magnitude of conductance at the Fermi level and the conduction spectra as a function of the energy of the incident electrons for the dumbbell-shaped dimer are in agreement with those reported in another theoretical study.\cite{PhysRevLett.103.206803} Since the dimers are not connected by the [2+2] four-membered rings in a direction parallel to the electrode and the 1D C$_{60}$ polymers act as an insulator, the conductance of the dimers is low around the Fermi level. This implies that the [2+2] four-membered rings play an important role in the metallic characteristics of the C$_{60}$ polymers and the very bright C$_{60}$ molecule in the experimental STM image\cite{nakaya} is presumably linked by the [2+2] four-membered rings and the peanut-shaped connections as explained in the preceding section. The difference in energy between the peaks of the spectrum around the Fermi level corresponds to the energy gap between the HOMO and the LUMO of the isolated dimers. The peak at $E_F$ + 0.25 eV is attributed to the triply degenerate LUMO ($t_{u1}$) of a C$_{60}$ molecule, where $E_F$ is the Fermi level. The conductance of the peanut-shaped dimer is higher than that of the dumbbell-shaped dimer below the Fermi level and the peak of the conductance spectrum induced by the $t_{u1}$ orbitals of the dumbbell-shaped dimer is clearer than that of the peanut-shaped dimer.

\begin{figure*}
\includegraphics{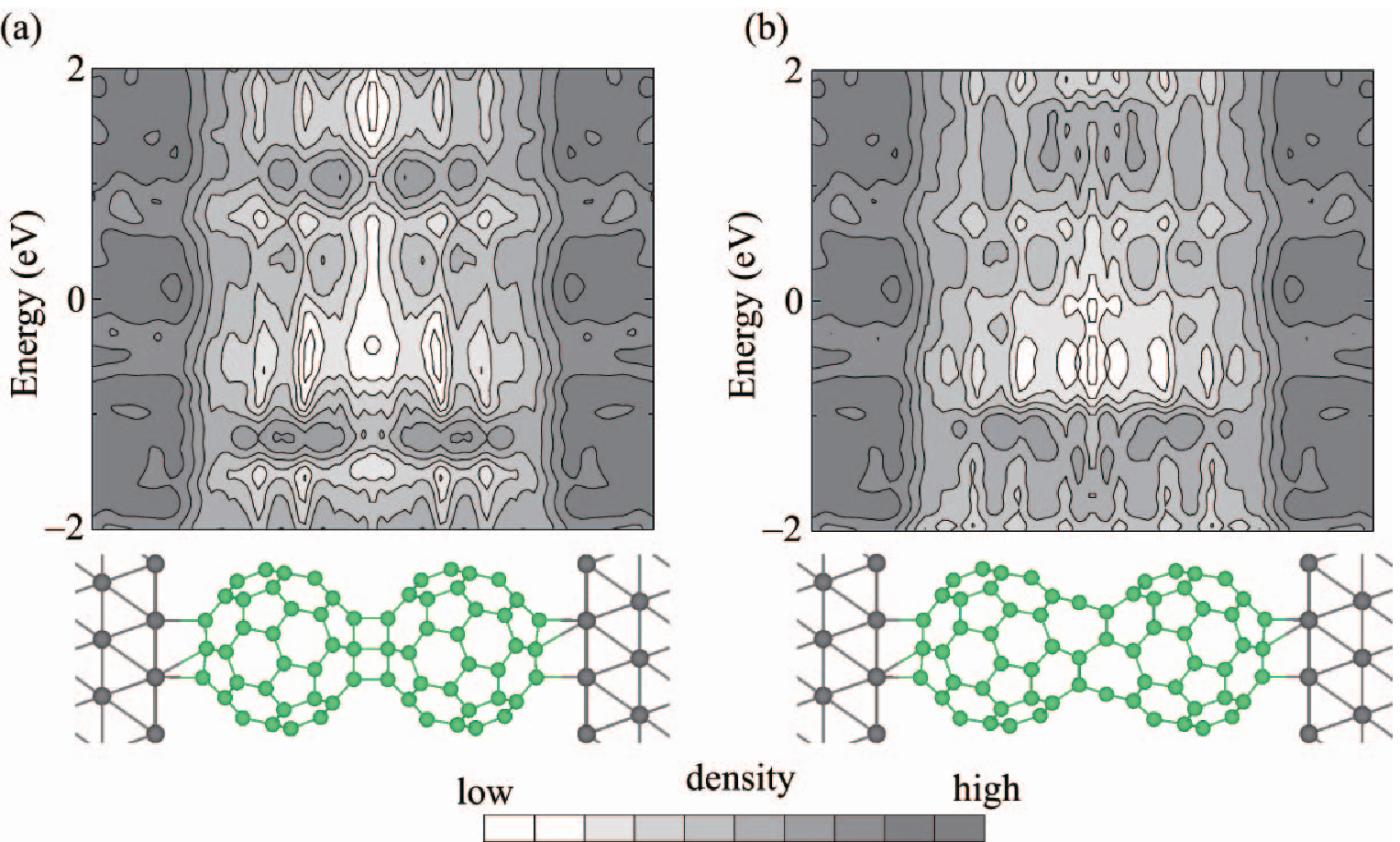}
\caption{(color online) Distributions of LDOS integrated on plane parallel to dimer as functions of relative energy from Fermi level. Zero energy is chosen as the Fermi level. Each contour represents twice or half the density of the adjacent contour lines, and the lowest contour is 2.56 $\times 10^{-4}$ {\it e}/eV/bohr. The atomic configurations below the graph are visual guides.}
\label{fig:5}
\end{figure*}

To examine the contribution of the peanut-shaped connection to the electron-transport properties in more detail, Fig.~\ref{fig:5} shows the LDOS of the dimers, which have been plotted by integrating them along the plane parallel to the electrode surface, $\rho(z,E)=\int |\psi(\vecvar{r},E)|^2 d\vecvar{r}_{||}$, where $\vecvar{r}=(x,y,z)$, $\psi$ is the wave function and $E$ is the energy of the states. The LDOS at $E_F-0.2$ eV around the molecule junction of the peanut-shaped dimer is larger than that of the dumbbell-shaped dimer. The energetically discrete $t_{u1}$ orbitals of the C$_{60}$ molecules contributing the electron transport change to the broadened states because of the formation of the $sp^2$-like connection at the molecule junction. The states in the electrode can easily penetrate into the peanut-shaped dimer and the conductance spectrum in Fig.~\ref{fig:4} becomes obscure. Although our computational models do not directly correspond to the tip and surface system of STS, the absence of clear peaks of the $t_{u1}$ orbitals at the center of the polymerized cluster in the STS spectrum is related to the smooth behavior of the conductance spectrum induced by the $sp^2$-like peanut-shaped connections.

\section{Summary}
\label{sec:summ}
The atomic configuration of the EB irradiated C$_{60}$ polymers has been proposed to explain the generation of the conductivity in the C$_{60}$ film through first-principles calculations. Although the polymer composed of the [2+2] four-membered rings and dumbbell-shaped interlayer connections is a semiconductor with a narrow band gap, the polymer changes to exhibit metallic characteristics by forming a peanut-shaped interlayer connection. The calculations for electron transport for the dumbbell-shaped and peanut-shaped dimers revealed that the higher conductance of the peanut-shaped dimer below the Fermi level than that of the dumbbell-shaped dimer and the low peak of the conductance spectrum induced by the $t_{u1}$ orbitals of the peanut-shaped dimer are relevant to the energetically broadened LDOS at the molecule junction caused by the formation of the $sp^2$-bonded peanut-shaped connection and the metallic property of the EB-irradiated C$_{60}$ polymer. Rhombohedral film consisting of [2+2] four-membered rings, dumbbell-shaped and peanut-shaped connections is a possible structure for the metallic C$_{60}$ polymer observed by STM at the initial stage of EB irradiation, since the present structure could be formed without significant structural deformation from face-centered cubic C$_{60}$ bulk and agrees with the rhombohedral structure in the STM image. Although our computational models are limited to the simplified C$_{60}$ clusters because of the limitation of the computational resources, the large-scale first-principles calculations for the C$_{60}$ polymers could help to validate the correlation between the STS spectra and the present atomic configuration of the polymers and will be carried out in a future.

\section*{Acknowledgements}
The authors would like to thank Kikuji Hirose and Yoshitada Morikawa of Osaka University for fruitful discussion. This research was partially supported by Strategic Japanese-German Cooperative Program from Japan Science and Technology Agency and Deutsche Forschungsgemeinschaft, by a Grant-in-Aid for Scientific Research on Innovative Areas (Grant No. 22104007) from the Ministry of Education, Culture, Sports, Science and Technology, Japan. The numerical calculation was carried out using the computer facilities of the Institute for Solid State Physics at the University of Tokyo, the Research Center for Computational Science at the National Institute of Natural Science, Center for Computational Sciences at University of Tsukuba, and the Information Synergy Center at Tohoku University.

\end{document}